\documentclass[aps,prl,preprint,secnumarabic,amsppt,superscriptaddress,showpacs]{revtex4}

\usepackage{graphicx}
\usepackage{amsmath,amsfonts,amssymb}
\usepackage{epsfig}
\usepackage{wrapfig}
\usepackage{bbm}
\usepackage[usenames]{color}
\usepackage{array}
\usepackage{times}

\begin{document}

\title{Imaging topological edge states in silicon photonics}

\author{M. Hafezi \footnote{Correspondence should be addressed to hafezi@umd.edu}}
\author{S. Mittal}
\author{J. Fan}
\author{A. Migdall}
\author{J. M. Taylor}
\affiliation{Joint Quantum Institute, National Institute of Standards and Technology / University of Maryland, College Park MD 20742}

%\setstretch{2}
\begin{abstract}

\textbf{Topological features -- global properties not discernible locally -- emerge in systems from liquid crystals to magnets to fractional quantum Hall systems.  Deeper understanding of the role of topology in physics has led to a new class of matter: topologically-ordered systems.  The best known examples are quantum Hall effects, where insensitivity to local properties manifests itself as conductance through edge states that is insensitive to defects and disorder. Current research in engineering topological order primarily focuses on analogies to quantum Hall systems, where the required magnetic field is synthesized in non-magnetic systems. Here, we realize  synthetic magnetic fields for photons at room temperature, using linear Silicon photonics. We observe, for the first time, topological edge states of light in a two-dimensional system and show their robustness against intrinsic and introduced disorder. Our experiment demonstrates the feasibility of using  photonics to realize topological order in both the non-interacting and many-body regimes.}

\end{abstract}
\maketitle

Charged particles in two-dimensional structures with a magnetic field exhibit a remarkable variety of macroscopic quantum phenomena, including integer-\cite{Klitzing:1980} and fractional- quantum Hall \cite{tsui82}, quantum spin Hall effects \cite{Konig:2007}, and the theoretical prediction of the emergence of particles with fractional statistics, so-called anyons \cite{LesHouchesHall:BOOK}.  Despite the great success in electronic systems, advances in the experimental efforts have been hampered by stringent experimental requirements such as purity. Recently, neutral ultracold gases have been theoretically and experimentally studied for observation of these effects \cite{Cooper:2008p6565, Spielman:2011, Bloch:2011, Dalibard:2011}. Ultracold atomic systems are advantageous since they provide tools for in situ  control of most parameters describing quantum Hall systems, however, strong effective magnetic field remains elusive. In contrast, photons, alleviating many of these experimental difficulties, provide a new avenue for investigation of quantum Hall physics at room temperature. Arrays of coupled optical resonator provide a toolbox, in the context of quantum simulation,  to engineer several classes of Hamiltonians and allow the direct observation of the wave function.  Furthermore, such photonic systems might find application in optical devices such as filters, switches and delay lines by exploiting topological robustness \cite{Hafezi:2011delay}.  Here, we report the first implementation of a magnetic-like Hamiltonian in a two dimensional photonic system and present the direct observation of robust edge states --- the hallmark of topological order.

Many proposals for implementing magnetic-like Hamiltonians in photonic systems require an external field such as large magnetic field \cite{Haldane:2008,Wang:2009p16784}, strain \cite{Rechtsman:2012}, harmonic modulation \cite{Fan:vy} or optomechanical induced non-reciprocity  \cite{Hafezi:2011ui}.  However, it has been shown that using an external field is not necessary; instead, by using either a polarization scheme \cite {Umucalilar}, differential optical paths \cite{Hafezi:2011delay}, or bi-anisotropic metamaterials  \cite{Khanikaev}, one can achieve a magnetic-like Hamiltonian in direct analogy to a spin-orbit interaction in electronic systems \cite{Bernevig:2006p27579}. Topological states of light have also been explored in one-dimensional photonic systems \cite{Kraus:2012}. Moreover, such systems could have direct applications in silicon photonics \cite{Hafezi:2011delay,Fan:vy}. Specifically, we implement a synthetic gauge potential using an induced pseudo-spin-orbit interaction where a time-reversed pair of resonator modes -- clockwise and counter-clockwise circulation -- acts as a pseudo-spin. Due to the large size of the resonators (several tens of microns) such photonic implementation  of a magnetic-like Hamiltonian allows direct observation of the wave function via optical imaging. We observe three primary experimental observations: (1) light propagates  along the system edges where the boundaries are defined as either the magnetic domains or the physical edges, (2) the light propagation profile of the edge states remains  unchanged over a broad band as a signature of robustness against intrinsic disorder, (3) the edge state propagation is robust against an introduced disorder: the transport is not impeded even in the absence of a resonator on the edge.

\textbf{Experimental implementation}

For the experimental implementation, we fabricated a two-dimensional (2D) array of coupled optical ring resonators where the design of the waveguides in the array allows us to simulate a magnetic field for photons using  silicon-on-insulator (SOI) technology \cite{Little:1998,Melloni:2002,Sakai:2004,Dumon:2004,Vlasov:2004,Xu:2005p29862,Xia:2007}. High-Q ring resonators  ($Q>10^4$) were fabricated on a SOI wafer with a 220 nm thick layer of silicon  on top of a 2 $\mu$m thick buried oxide (BOX) layer that isolates the optical mode and prevents it from leaking to the substrate. The cross section of the waveguides, which forms the link and site resonators, was designed to be 510 nm $\times$ 220 nm, to assure single-mode propagation of the transverse-electric (TE) light (the electric field in the slab plane) at the telecom wavelength $(\approx$ 1.55 $\mu$m). The typical air gap for evanescent coupling between the site-resonators and the probing waveguides (link-resonators) was chosen to be 180 nm (200 nm). The 90$^{o}$ bending radius of the rounded rectangles was chosen to be 6 $\mu$m to have the bending loss negligibly small  \cite{Vlasov:2004}. The fabrication of silicon chips was performed through ePIXfab, and set-up by Leti-CEA and IMEC facilities. The masks were made using deep ultraviolet 193 nm photolithography and were etched in two steps of 70 nm/220 nm, for gratings/waveguides, respectively. The process was followed by thermal oxidation (10 nm) to reduce surface roughness. In our experimental setup,  grating couplers were used for input and output coupling to the device. Light scattered from the resonators was spatially imaged using a 25x microscope objective  and an InGaAs infrared camera (640 $\times$ 512 pixel grid with a 25 $\mu$m pitch), as shown in Fig.~1c. Such a setup allows us to measure the relative amount of  light scattered from each site \cite{Cooper:2010ua}.  Transmission through the device was measured using an optical vector analyzer (Luna technologies OVA 5000).

To describe the essence of the scheme, we consider a single plaquette of our lattice which consists  of four ``site-resonators'' and four ``link-resonators'' in the form of rounded rectangles,  as shown in Fig.~1a. The link- and site-resonators are coupled through directional couplers to each other, and therefore, photons circulating in one direction in the site-resonators only couple with each other and with photons circulating in the opposite direction in the link-resonators. The effective length of the link-resonators is chosen to be larger than that of the site-resonators by $2\eta$, so that the links and sites are resonant at different frequencies. Consequently, a photon resonant with the site-resonators spends substantially more time in the sites than in the links. We associate the clockwise photons in site-resonators  with the up component of a pseudo-spin.  By virtue of the time-reversal symmetry, the pseudo-spin down component (counter-clockwise photons in the site-resonators) is degenerate with the pseudo-spin up component. For the moment, we focus on the spin-up component. Depending on the positioning of the links, the photon acquires a different phase hopping forwards than backwards. In particular, the hopping process between sites 1 and 2 in Fig.~1a is described by: $\hat{a}^\dagger_2\hat{a}_1 e^{-i\phi_{12}}+\hat{a}^\dagger_2\hat{a}_1 e^{i\phi_{12}} $, where $\hat{a}_i$ is the creation operator of a photon at site $i$. The phase arises from an offset of the link waveguides from the symmetric point (defined as equal amounts of additional length above and below the directional coupler).  Specifically, the additional phase is given by the optical length $\phi_{12}=4\pi n x_{12}/\lambda$, where $n$ is the index of refraction, $x_{12}$ is the position shift of the link resonator, and $\lambda$ is the wavelength of the light. Note that  the additional length $\eta$ and position shifts away from the symmetric point are designed to keep the lengths of the directional couplers, the geometry of their coupling regions, and their coupling efficiencies invariant (Fig.~1a). Thus, the overall Hamiltonian describing photon hopping in the plaquette can be written:
\begin{equation}
-J\left[  \hat{a}^\dagger_2\hat{a}_1 e^{-i\phi_{12}}+\hat{a}^\dagger_3\hat{a}_2+\hat{a}^\dagger_4\hat{a}_3 e^{i\phi_{34}}+\hat{a}^\dagger_1\hat{a}_4\right]+h.c.,
\end{equation}
where $J$ is the tunneling rate  and the photon going counter-clockwise around the plaquette acquires a $2\pi\alpha$ phase where $\alpha=2 n(x_{34}-x_{12} )/\lambda$. If the phase per plaquette is uniform over a region, the photonic dynamics are equivalent to those of charged particles in  a uniform perpendicular magnetic field \cite{Hafezi:2011delay}. Such a system is predicted to exhibit edge states at the boundaries of that region \cite{Laughlin:1981p37080,Halperin:1982}. In a photonic system, such edge states can be excited by driving the system at specific frequency bands.

To verify that the expected edge physics arises entirely from our synthetic gauge field, we first designed a phase slip between $10\times4$ stripes, as shown in Fig.~1b. This results in having magnetic domains that are entirely due to passive, and controlled, interference effects.  The resulting edge states of the system then follow along the edge of the magnetic domains induced by this phase slip (Fig.~1b), rather than the physical edge of the system (see Supplementary Information). The effective uniform magnetic field in the stripe is given by ($\alpha \approx 0.15 $). The dispersion of the system is shown in Fig.~2a, where the edge state band are shown between magnetic bulk bands. The light is coupled to the 2D ring resonators using a bent waveguide at the two bottom corners, as shown in Fig.~1b. Depending on the pumping direction, the two different pseudo-spin components can be excited, e.g. coupling light the system at port 1 (2), pumps the system in the spin-up (spin-down) component.

\textbf{Results}

As a demonstration of the scheme, we measured the transmission spectrum of the 2D system through various ports and compared it with our simulation, as shown in Fig.~2.  To do this, we first characterized different system parameters using simpler devices such as a notch filter (single resonator coupled to a waveguide), and an add/drop filter (single resonator coupled to two waveguides)   fabricated on the same chip to allow for calibration and characterization of the waveguides and resonators (see Supplementary Information). We estimate the probing waveguide-resonator coupling rate ($\kappa_{\rm{\rm{ex}}} \approx15$ GHz), the intrinsic loss ($\kappa_{\rm{\rm{in}}}\approx 1$ GHz) and the tunneling rate between site-resonators ($J \approx16$ GHz), where all measurements are within 2 nm of the center wavelength of $1539$ nm. Given these parameters, we simulate a $10\times10$ lattice using the transfer matrix formalism (Supplementary Information), as shown in Fig.~2a. We have also considered a random onsite impurity  shift of the resonance frequency with a standard deviation of $~0.8  J$.  In a lossless system,  the transmission spectrum for the spin-down ($T_{12}$) and spin-up ($T_{34}$) should be identical, although they may take different paths. However, the presence of loss breaks this symmetry. We observe a qualitative agreement between the simulation and experiment (Fig.~2b,c).

A crucial point which allows us  to implement the magnetic Hamiltonian in our photonic system is the weakness of backscattering. In principle, backscattering in the waveguide and resonators can flip the pseudo-spin and destroys the desired behavior. To address this, we used directional couplers to reduce the effect of backscattering in the coupling regions. However, surface roughness of the waveguides still leads to some backscattering, which can accumulate in our large 2D system. Thus,  the pseudo-spin can flip from up to down (i.e., clockwise to counter-clockwise) and leave the system by the opposite channel (e.g. when the system is pumped with pseudo spin up from port 1, the flipped spin exits the system at port 4). We confirm that the level of backscattering, as characterized by $T_{14}$ in Figs. 2b-c, is small, as predicted previously \cite{Morichetti:2010p45154}.

To demonstrate  the presence of edge states in our system, we image the propagation of light in various devices. We start  by a system with magnetic domains, as shown in Fig.~1b. In the presence of a magnetic field, one expects edge states to appear when the system is pumped at certain frequencies \cite{Hafezi:2011delay}. According to the dispersion simulation (Fig.~2a), we expect the absence of bulk states and presence of edge states to occur at the frequency interval $\omega/J=2.5\pm0.5$. Fig.~2d shows such an edge state starting at the input, routing around the $10\times4$ stripe edge, and leaving the system at the output. In the experiment, we observe a similar behavior, as shown in Fig.~2e. Note that compared to the simulation, the edge state in the experiment undergoes higher loss and attenuates more rapidly before getting to the other side of the 2D system. The remarkable feature of such edge states is that they route around the boundary that is defined by our synthetic magnetic field, rather the the physical edge of our system. When the system is pumped at a different frequency, bulk states are excited which do not have a particular shape, as shown in Figs.~2f,g. States propagating in the bulk are more susceptible to the frequency mismatch of the resonators, and therefore, one should not expect that the spatial profile to match the numerical simulation. More importantly, the edge state profile does not significantly change over a band broader than 5 GHz. In contrast,  the profile of the bulk states changes dramatically when the frequency is changed even by 0.2 GHz. We attribute this remarkable difference to the topological protection of the edge states.

Next, we imaged the edge state for a system which was designed without a phase slip, i.e., the magnetic field is uniform over the entire system. Fig.~\ref{fig:uniform_magnetic_field} shows the light  propagation along the short (Fig.~3a,c) and the long (Fig.~3b,d) edges. The light is launched at a specific frequency band $\omega/J=1.7\pm0.6 (-1.7\pm0.6)$, corresponding to the short (long) edge excitation. The  physical transverse width of the edge state is about 1-2 resonators as observed both in the experiment and the simulation. The width is slightly greater in the experiment than in the numerical simulation, due to the presence of intrinsic disorder in fabrication which has been ignored in the simulation. Due to the topological robustness of the system, in the long edge band, the light takes two sharp corners, as shown  in Fig.~\ref{fig:uniform_magnetic_field}  b-d, and leaves the system at the output port. Again, similar to the previous case, we observe that the edge state profile is robust over a band (10 GHz), broader than the bulk states (0.2 GHz), due to the topological protection of the edge states. Note that if the system were isolated (i.e., in the absence of input/output ports), the edge states can circulate around the entire system. Once the system is coupled to input/output ports, depending on the excitation frequency, either the short or the long edge acquires higher coupling and the resonators exhibit higher light scattering.

Lastly, to confirm that the edge state is robust against an introduced disorder, we fabricated a 10x10 array with a missing resonator on one of the edges, as shown in Fig.~\ref{fig:defect}a. Due to topological robustness, the edge state is expected to bypass the impurity and the transmission should not be impeded. Launching light at the short-edge band (over 15 GHz), we observed that the light enters at the bottom row from the left input corner, routed tightly around the disorder  --- in this extreme case,  an entirely missing resonator ---  and travelled to the output port, without entering into the bulk (Fig.~\ref{fig:defect}b), in good agreement with the simulation (Fig.~\ref{fig:defect}c).  For the simulation, we have used the parameters that were independently measured using single ring devices. The residual presence of light at the missing resonator location in Fig.~\ref{fig:defect}b, is due to the background noise.

\textbf{Conclusion}

We have demonstrated the presence of robust edge states in an engineered  two-dimensional photonic system. The  silicon-on-insulator technology allowed us to measure both the  transport properties and the spatial wave function of  such states. This platform opens the door to study different types of magnetic fields and topological orders with photons in the non-interacting regime, due to the resonant enhancement. Moreover, intriguing avenues can be seen for exploring many-body physics by integrating strong nonlinearity, such as one mediated by quantum-dots \cite{Waks:2002, Srinivasan:2007p4061,Englund:2007,AndreiFaraon:2008p45019} or Rydberg atomic ensembles \cite{Dudin:2012hm,Peyronel:2012}.
\\
\\
\textbf{Acknowledgments:} We thank  J. Chen, K. Srinivasan, M. Lukin, A. Melloni, M. Fournier, G. Solomon, and E. Waks  for stimulating discussions and experimental help. M.H. thanks the Institute for Quantum Optics and Quantum Information, Innsbruck for hospitality. This research was supported by the U.S. Army Research Office Multidisciplinary University Research Initiative award W911NF0910406, and the NSF through the Physics Frontier Center at the Joint Quantum Institute.
\\
\\
\textbf{Disclaimer:} Certain commercial equipment, instruments or materials are identified in this paper to foster understanding. Such identification does not imply recommendation or endorsement by the National Institute of Standards and Technology, nor does it imply that the materials or equipment are necessarily the best available for the purpose.
\\
\textbf{Authors contribution:} M.H. conceived the experiment. M.H., J.F., A.M. and J.M.T designed the chips. M.H., S.M. and J.F. carried out the experiment and analyzed the data. M.H. and J.M.T. wrote the manuscript. All authors contributed considerably.
The authors declare no competing financial interests. Correspondence and requests for materials should be addressed to M.H. (hafezi@umd.edu).

\newpage
\begin{figure}
\includegraphics[width=\textwidth]{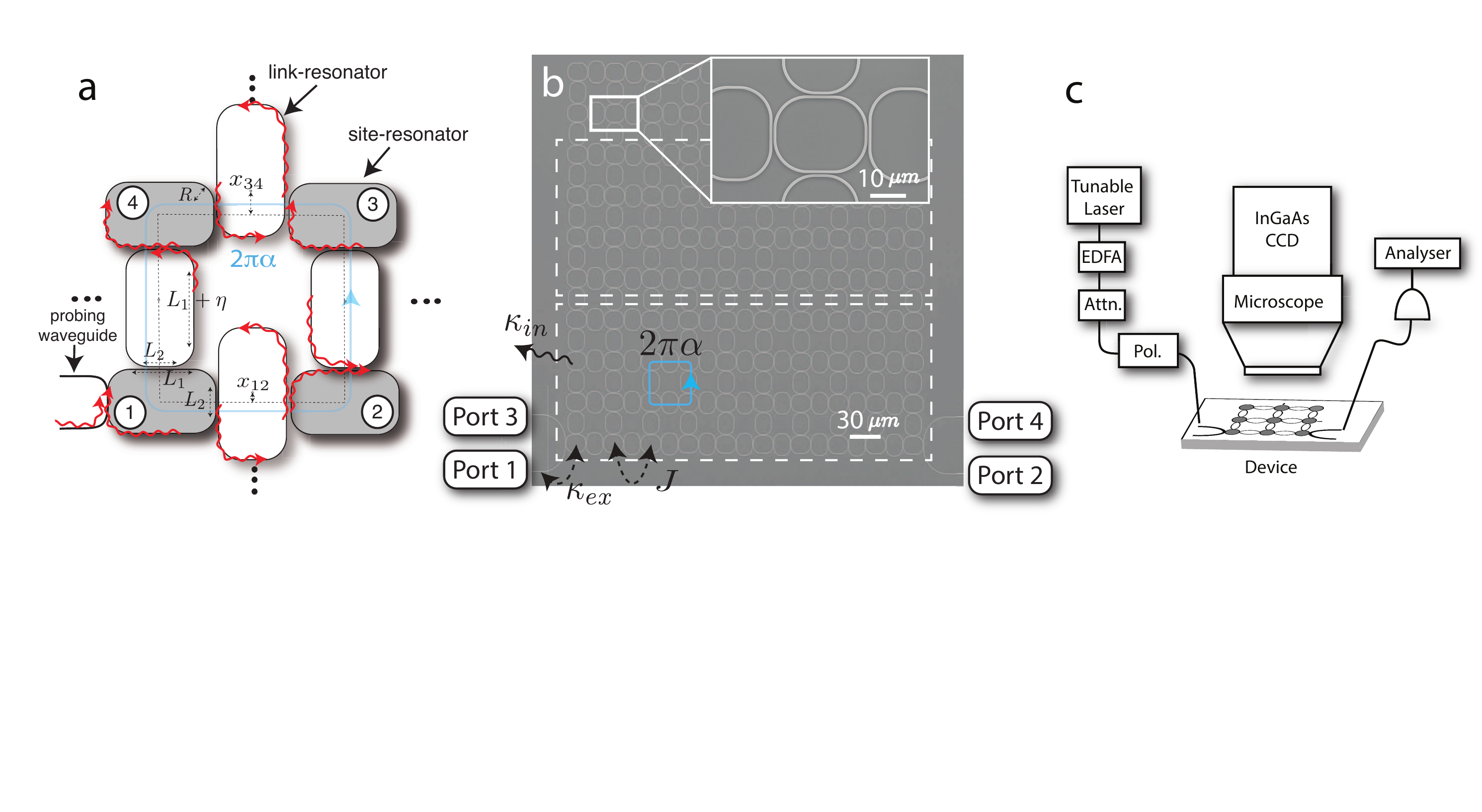}\caption{\textbf{Experimental setup:} (a) A  single plaquette  consisting of 4 link resonators and 4 site resonators: gray (white) rounded rectangles represent  site- (link-) resonators.  These two types of resonators differ due to an extra length $(2\eta)$ in the link-resonators.  Moreover, due to the vertical shift of the link-resonators, a photon acquires a non-zero phase when it hops between resonators (1,2) and (3,4). Therefore, a photon going counter-clockwise (clockwise) around the plaquette acquires a $2\pi\alpha (-2\pi\alpha)$ phase. (b) the Scanning electron microscope (SEM) image of the device. The stripes with uniform magnetic field are delineated with white dashed lines. (c) Schematic of the experimental setup.
\label{fig.1} }\end{figure}

\newpage
\begin{figure}
\includegraphics[width=\textwidth]{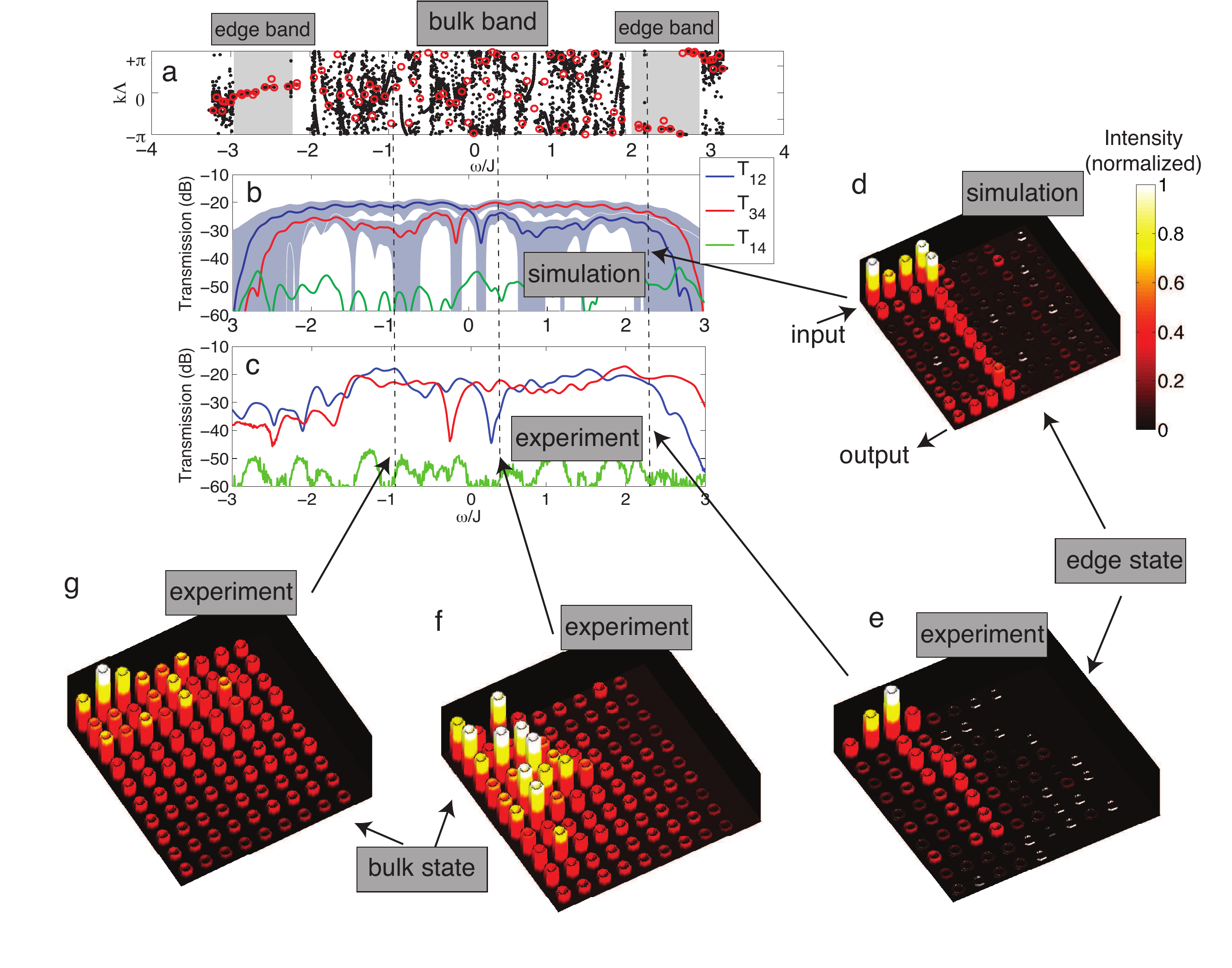}\caption{\label{fig.2} \textbf{Edge states around a magnetic domain:} (a) Simulated dispersion of the system: k$\Lambda$ is the relative phase between two adjacent resonators on the edge while $\omega$ is the relative detuning of  frequency with respect to the band center, in units of the tunneling rate. Red dots represent the dispersion of our system, black dots represent simulation of a longer system (10$\times$400), to better distinguish the bulk band  from the edge band.  (b) Simulated transmission of a 10x10 lattice. $T_{ij}$  is the transmittance  between port $i$ and $j$, as shown in Fig.1b. $T_{14}$ measures the backscattered light. Simulation parameters are estimated from the experiment. (c) Measured transmission spectrum. (d) Simulated scattered light from a 2D array of the couple resonators, when the system is pumped at the edge state band (e) Image of the edge state; the system is pumped at the frequencies corresponding to edge state band. (e,g) Image of bulk states; the state is pumped at frequencies that are in the bulk state band.}\end{figure}

\newpage
\begin{figure}
\includegraphics[width=.8\textwidth]{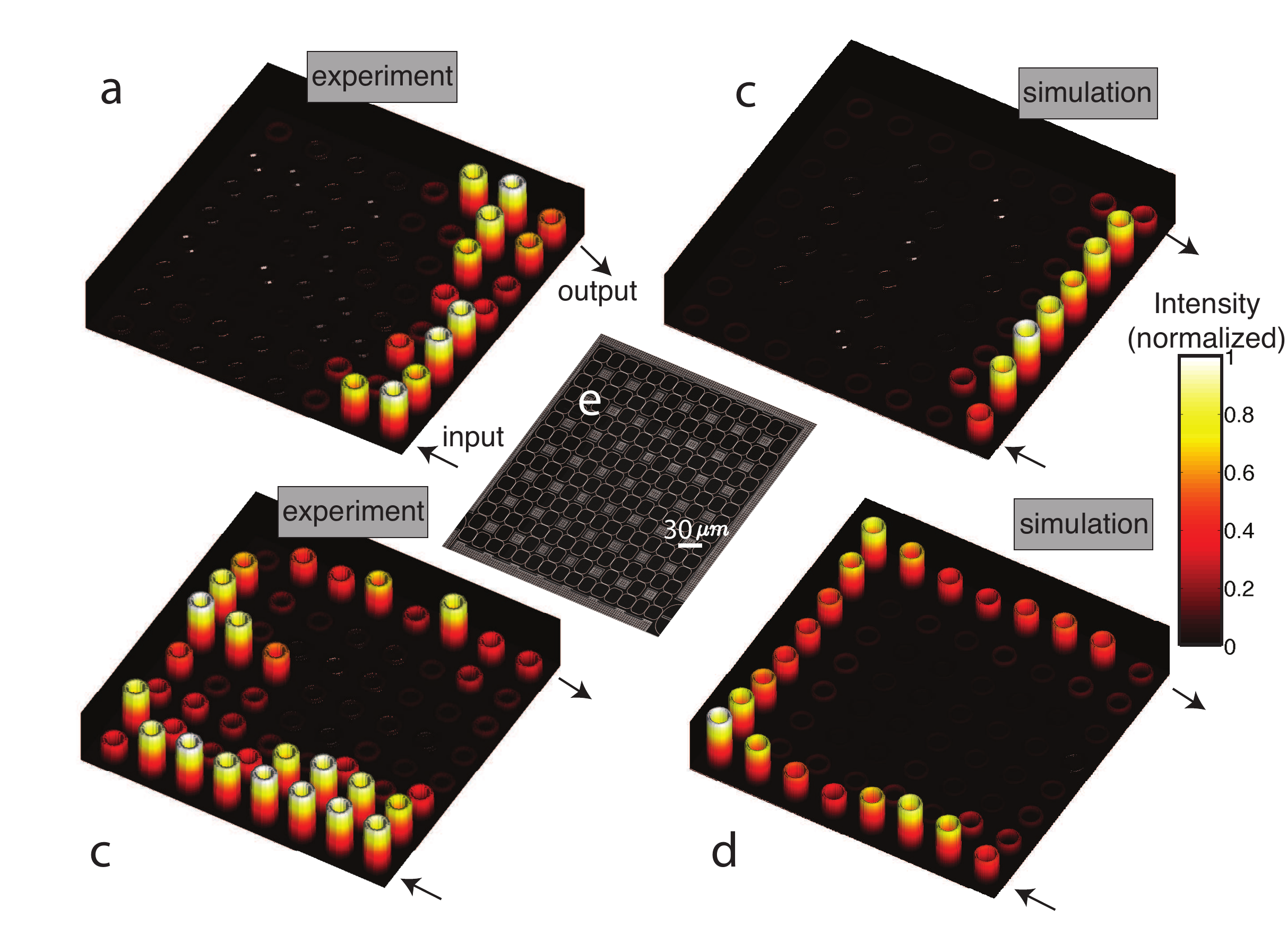}\caption{ \textbf{Edge state propagation in a homogenous magnetic field (8x8 array):} The light enters from one corner and exits from the other corner. The experiment shows that depending on the input frequency, the light takes the short edge (a) or the long edge (b).   The experimental results (a-b) are in good agreement with the simulation results (c-d).  The simulation parameters are $(\kappa_{\rm{ex}},\kappa_{\rm{\rm{in}}},J)=(31,.57,26)$GHz which are extracted from experimental measurement of simpler devices. (e) An SEM image of the system. \label{fig:uniform_magnetic_field} }\end{figure}

\newpage
\begin{figure}
\includegraphics[width=.8 \textwidth]{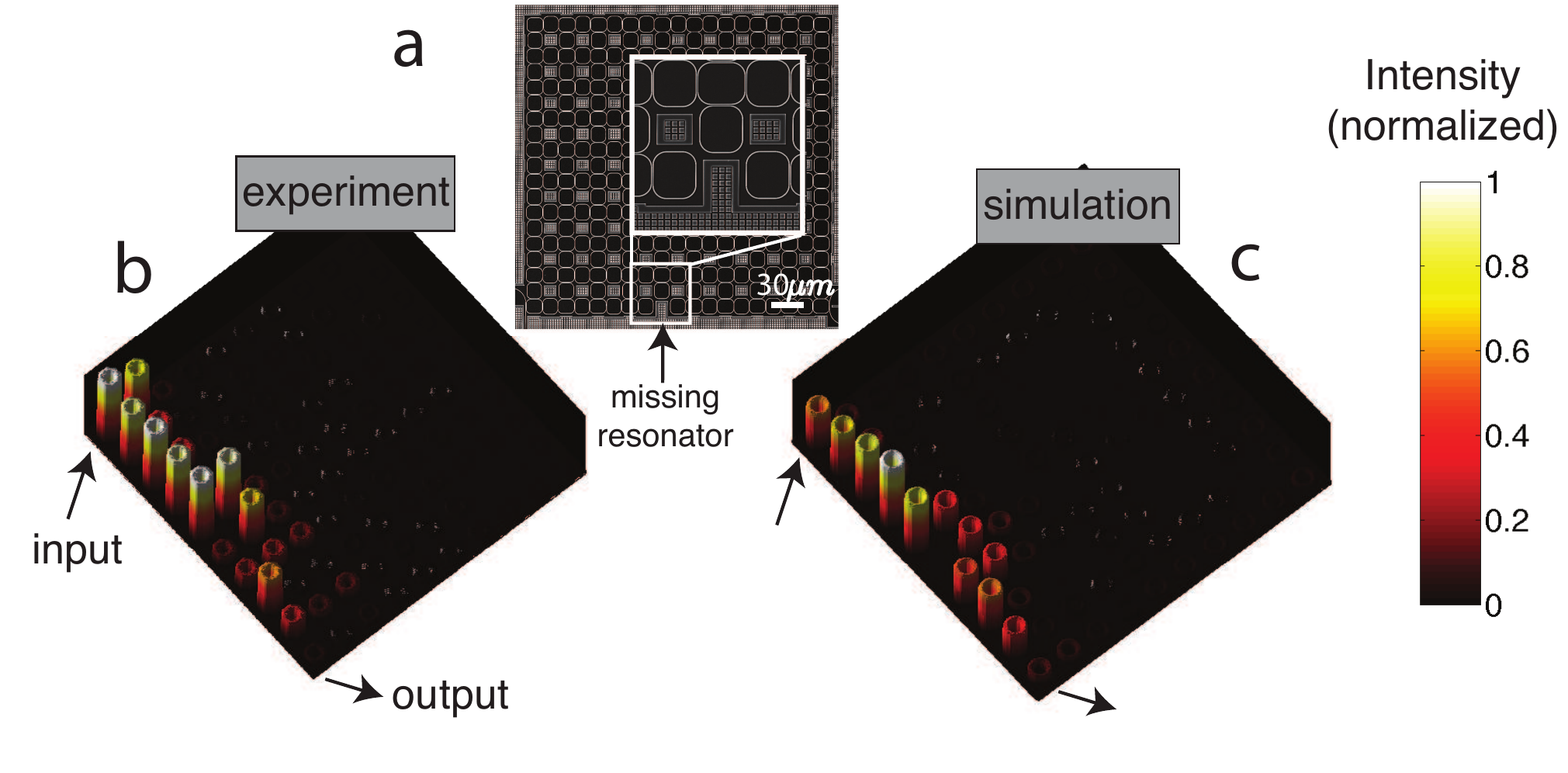}\caption{ \textbf{Edge state protection against a defect:} A resonator was intentionally removed from the array, as shown in the SEM image (a). Topological protection is observed in the experiment (b) as light propagating along the edge routes around the defect, in agreement with the simulation (c). The parameters for the simulation are the same as Fig.~\ref{fig:uniform_magnetic_field}. \label{fig:defect}} \end{figure}

\end{document}

\newpage
\section*{Supplementary Information for Imaging topological edge states in Silicon photonics }
\section*{ S1. Inducing a magnetic-like phase in coupled optical resonators}

Here, we show that the dynamics of two ring resonators coupled
through a middle off-resonant ring can be effectively described as two
resonators coupled with a ``hopping phase''. We evaluate transport
properties in both cases and show that they are identical, near
the resonant frequency of the side resonators.

\begin{figure}[h]
\includegraphics{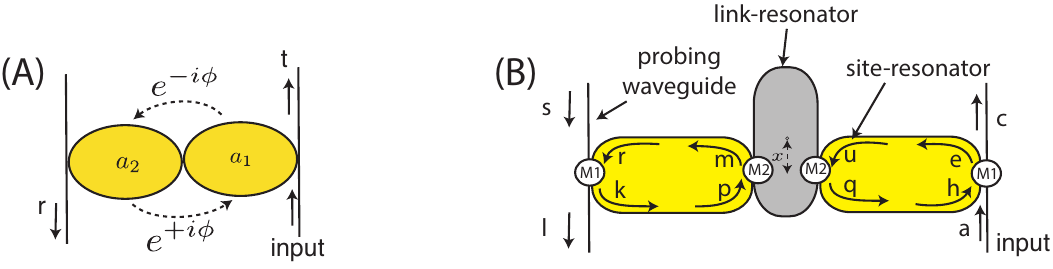}\caption{(a) Two ring resonator coupled with a hopping phase $(\phi)$
(b) Two ring resonators coupled through an off-resonant middle ring.
These two systems are equivalent around the resonant frequency of
side resonators.\label{fig:three-rings}}
\end{figure}

First, we derive the transmission and reflection coefficients of two
rings coupled with a hopping phase. The Hamiltonian describing
such system can be written as:

\begin{equation}
H=-J\hat{a}_{2}^{\dagger}\hat{a}_{1}e^{-i\phi}-J\hat{a}_{1}^{\dagger}\hat{a}_{2}e^{+i\phi},
\end{equation}
where $J$ is the tunneling rate and $\phi$ is the hopping
phase, as depicted in Fig.~(1A). Note that by hopping phase, we mean
a Hamiltonian of the kind written in Eq.(1). This should not be confused with optical non-reciprocity which requires an external field, cf.
Refs. \cite{Hafezi:2011ui,Fan:vy}.

We assume the resonators are single mode. Therefore, using coupled
mode theory (input-output formalism), we can write the dynamics of
the field inside the resonators as:

\begin{equation}
\frac{d}{dt}\left(\begin{array}{c}
a_{1}\\
a_{2}
\end{array}\right)=\left(\begin{array}{cc}
-\text{\ensuremath{\kappa_{\rm{\rm{in}}}}}-\ensuremath{\kappa_{\rm{ex}}} & iJe^{+I\phi}\\
iJe^{-I\phi} & -\text{\ensuremath{\kappa_{\rm{in}}}}-\ensuremath{\kappa_{\rm{ex}}}
\end{array}\right)\left(\begin{array}{c}
a_{1}\\
a_{2}
\end{array}\right)-\sqrt{2\text{\ensuremath{\kappa_{\rm{ex}}}}}\left(\begin{array}{c}
\mathcal{E}_{\rm{in}}\\
0
\end{array}\right),
\end{equation}
where $\kappa_{\rm{ex}}$ is the coupling rate between the probing-waveguide
and the side resonators and $\kappa_{\rm{in}}$ is the field decay rate
to undesired modes. We assume a monochromatic input field at resonator one with amplitude $\mathcal{E}_{\rm{in}}$ and detuned by $\omega$
from the resonance, as shown in Fig.(\ref{fig:three-rings}a). The output field of the
resonators will be: $a_{2}^{\rm{out}}=\sqrt{2\kappa_{\rm{ex}}}a_{2}$, $a_{1}^{\rm{out}}=1+\sqrt{2\kappa_{\rm{ex}}}a_{1}$.
Consequently, the transmission and reflection coefficients, as defined
in Fig.~\ref{fig:three-rings}, are given by:

\begin{eqnarray}
r_{\rm{SM}} & = & \frac{\sqrt{2\kappa_{\rm{ex}}}a_{2}}{\mathcal{E}_{\rm{in}}}=-\frac{2ie^{-i\phi}J\text{\ensuremath{\kappa_{\rm{ex}}}}}{J^{2}+(i\omega-\text{\ensuremath{\kappa_{\rm{ex}}}}-\text{\ensuremath{\kappa_{\rm{in}}}})^{2}}\label{eq:dropSM}\\
t_{\rm{SM}} & = & \frac{\sqrt{2\kappa_{\rm{ex}}}a_{1}+1}{\mathcal{E}_{\rm{in}}}=1+\frac{2\text{\ensuremath{\kappa_{\rm{ex}}}}(+i\omega-\text{\ensuremath{\kappa_{\rm{ex}}}}-\text{\ensuremath{\kappa_{\rm{in}}}})}{J^{2}+(+i\omega-\text{\ensuremath{\kappa_{\rm{ex}}}}-\text{\ensuremath{\kappa_{\rm{in}}}})^{2}}.\nonumber
\end{eqnarray}

Now, we consider two ring resonators that are coupled through a middle
off-resonant ring, as shown in Fig.(\ref{fig:three-rings}b). We use the transfer matrix
formalism to derive the transmission of the system. The transfer matrix
for the waveguide-resonator and resonator-resonator coupling regions,
respectively, are given by:

\begin{eqnarray*}
M_{1} & = & \frac{1}{t_{1}}\left(\begin{array}{cc}
-r_{1}^{2}+t_{1}^{2} & r_{1}\\
-r_{1} & 1
\end{array}\right)\\
M_{2} & = & \frac{1}{t_{2}}\left(\begin{array}{cc}
-r_{2}^{2}+t_{2}^{2} & r_{2}\\
-r_{2} & 1
\end{array}\right),
\end{eqnarray*}
where $t_{\rm{i}}$ and $r_{\rm{i}}$ are the transmission and reflection coefficients
of the coupling regions. Therefore, the probe waveguide-resonator
coupling can be written as:

\begin{equation}
\left(\begin{array}{c}
e\\
h
\end{array}\right)=M_{1}\left(\begin{array}{c}
a\\
c
\end{array}\right)\text{ }\text{ },\text{ }\text{ }\left(\begin{array}{c}
l\\
s
\end{array}\right)=M_{1}\left(\begin{array}{c}
r\\
k
\end{array}\right).
\end{equation}
The input is at the right resonator, as shown Fig.(\ref{fig:three-rings}b), so we can
replace $a=1,s=0$. We assume that the length of the side resonators
(middle resonator) are $L\,(L+\eta)$, so that the middle resonator
is off-resonant with the other two. We assume the propagation
constant is $\beta=2\pi n/\lambda$, where $n$ is the index of refraction
and $\lambda$ is the wavelength, and the absorption constant is $\alpha'$.
The free propagation inside the side resonators will be given by:

\begin{equation}
\left(\begin{array}{c}
u\\
q
\end{array}\right)=\left(\begin{array}{cc}
e^{i\text{\ensuremath{\beta}L}/2-\text{\ensuremath{\alpha'}L}/2} & 0\\
0 & e^{-i\text{\ensuremath{\beta}L}/2+\text{\ensuremath{\alpha'}L}/2}
\end{array}\right)\left(\begin{array}{c}
e\\
h
\end{array}\right)\text{ },\text{ }\text{ }\left(\begin{array}{c}
r\\
k
\end{array}\right)=\left(\begin{array}{cc}
e^{i\text{\ensuremath{\beta}L}/2-\text{\ensuremath{\alpha'}L}/2} & 0\\
0 & e^{-i\text{\ensuremath{\beta}L}/2+\text{\ensuremath{\alpha'}L}/2}
\end{array}\right)\left(\begin{array}{c}
m\\
p
\end{array}\right)
\end{equation}
and the propagation inside the middle rings is given by:

\begin{equation}
\left(\begin{array}{c}
m\\
p
\end{array}\right)=M_{2}\left(\begin{array}{cc}
e^{i\text{ }\text{\ensuremath{\beta}L}/2+i\beta\eta-i2\text{\ensuremath{\beta}x}-\text{\ensuremath{\alpha'}L}/2} & 0\\
0 & e^{-i\text{ }\text{\ensuremath{\beta}L}/2-\text{ }i\beta\eta-i2\text{\ensuremath{\beta}x}+\text{\ensuremath{\alpha'}L}/2}
\end{array}\right)M_{2}\left(\begin{array}{c}
u\\
q
\end{array}\right).
\end{equation}
Using the above equations, we can obtain the transmission coefficients.
We are interested in the limit where the coupling loss can be ignored
(i.e. $|t_{\rm{i}}|^{2}+|r_{\rm{i}}|^{2}=1$) and the junctions are highly reflective.
In other words: $r_{\rm{i}}\to\sqrt{1-\epsilon_{\rm{i}}^{2}},\text{}t_{\rm{i}}\to i\epsilon_{\rm{i}}$,
where $\epsilon_{\rm{i}}\ll1$. The regime of interest is near the
resonant frequency of the side resonators, and is much smaller than the free
spectral range (FSR), so we consider $\beta L\ll1$. Since the propagation
loss over a typical distance in our experiment is not large, we take
$\alpha'L\ll1$. Keeping terms to 2nd order in $\epsilon_{\rm{i}}^{2},\beta L,\alpha'L$,
both in the numerator and the denominator, we find that the  field
in the drop channel can be simplified as:

\begin{equation}
r_{\rm{TM}}=\frac{2e^{-i\text{\ensuremath{\phi}}}\epsilon_{1}^{2}\epsilon_{2}^{2}}{2\left(2\text{\ensuremath{\alpha'}L}-2i\text{\ensuremath{\beta}L}+\epsilon_{1}^{2}\right)\epsilon_{2}^{2}\text{\ensuremath{\cos}}(\beta\eta)-i\left(\left(2\text{\ensuremath{\alpha'}L}-2i\text{\ensuremath{\beta}L}+\epsilon_{1}^{2}\right)^{2}+\epsilon_{2}^{4}\right)\text{sin}(\beta\eta)}.
\end{equation}

To compare this expression with the one obtained from the
single-mode approximation (Eq.(\ref{eq:dropSM})), we use the following substitutions:

\begin{equation}
\text{\ensuremath{\epsilon_{1}}}^{2}\to\frac{4\pi\text{\ensuremath{\kappa_{\rm{ex}}}}}{\text{FSR}}\text{ }\text{ },\text{ }\text{ }\text{\ensuremath{\epsilon_{2}}}^{2}\to\frac{4\pi J}{\text{FSR}}\text{ }\text{ },\text{ }\text{ }\text{\ensuremath{\alpha'}L}\to\frac{2\pi\text{\ensuremath{\kappa_{\rm{in}}}}}{\text{FSR}}\text{ }\text{ },\text{ }\text{ }\text{\ensuremath{\beta}L}\to2\pi\frac{\omega}{\text{FSR}}\text{ },\text{ }\text{ }\text{\ensuremath{\beta}x}\to\phi
\end{equation}
and we have

\begin{equation}
r_{\rm{TM}}=\frac{2e^{-i\text{\ensuremath{\phi}}}J\text{\ensuremath{\kappa_{\rm{ex}}}}}{2J(\text{\ensuremath{\kappa_{\rm{ex}}}}+\text{\ensuremath{\kappa_{\rm{in}}}}-i\omega)\cos(\beta\eta)-i\left(J^{2}+(\text{\ensuremath{\kappa_{\rm{ex}}}}+\text{\ensuremath{\kappa_{\rm{in}}}}-i\omega)^{2}\right)\sin(\beta\eta)}.
\end{equation}
We immediately see that if $\beta\eta=3\pi/2$, this expression reduces
to Eq.(\ref{eq:dropSM}). This means that if the middle
ring is anti-resonant with the side rings, the three rings can be
effectively described by two resonators coupled with a hopping
phase. If we have $\beta\eta=\pi/2$, the two models are again the same,
only the sign of the tunneling is reversed $(J\rightarrow-J)$. When the
middle resonator deviates from the anti-resonant condition $(\beta\eta=\pi/2,3\pi/2,...)$,
the system again can be effectively described  by two resonators with
a hopping phase, however, the effective tunneling is $J_{\rm{eff}}\rightarrow J/\sin(\beta\eta)$
and system is shifted in frequency by $\omega\rightarrow\omega-J\cot(\beta\eta)$.

\begin{figure}
\includegraphics[width=.8\textwidth]{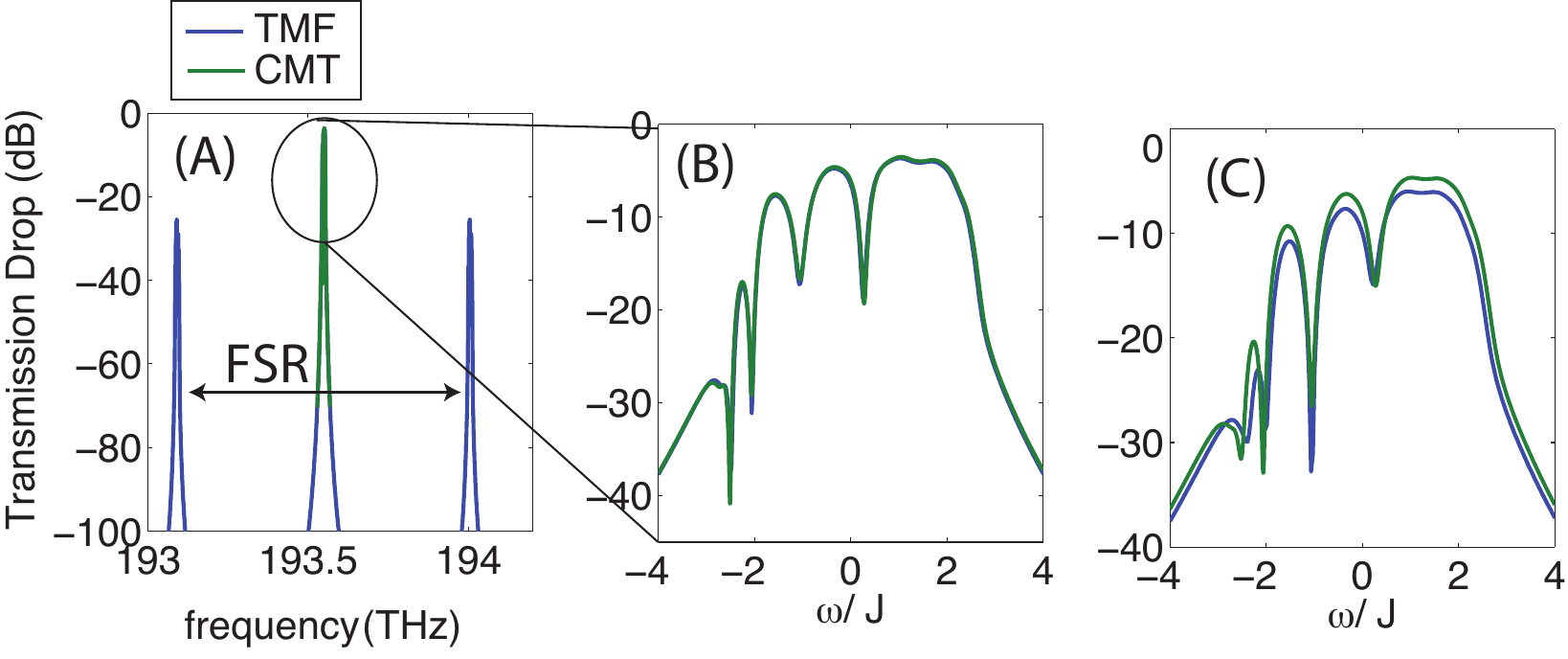}\caption{(a) Comparison between the coupled mode theory (CMT) and the transfer matrix formalism  (TMF) for a 4x4 array. (A) Both approaches agree near the resonant frequency of the site resonators. The side peaks are resonances of the link resonators. (B) shows a zoom in of (A) where the frequency is scaled by the tunneling rate J. The parameter are chosen to be $(\kappa_{\rm{ex}},\kappa_{\rm{in}},J)=(3,0.2,2)$ GHz. The remarkable agreement between both formalisms makes the difference invisible in the plot. (C) Same comparison for stronger couplings:  $(\kappa_{\rm{ex}},\kappa_{\rm{in}},J)=(20,2,10)$ GHz, two approaches slightly deviate from each other. For all the simulations, we have considered  a ring length of 80.3 $\mu$m so that the resonant wavelength coincides with $\lambda_0=1550$ nm. We have chosen $\eta=x=\lambda_0/ (\alpha n_{\rm{eff}})$, $\alpha=1/4$ and the effective (group) index is $n_{\rm{eff}}=2.47 (n_g=4.7)$.  \label{fig:CMT}}
\end{figure}

Similarly, for the 2D array, one can use the Hamiltonian description  (coupled mode theory) or the transfer matrix formalism. As discussed above, as long as the relevant bandwidths are smaller than the free spectral range (i.e., $J,\kappa_{\rm{ex}},\kappa_{\rm{in}}\ll $ FSR), the results of both formalisms coincide, as shown in Fig.~\ref{fig:CMT}.  Note that for the transfer matrix simulation, we have also included dispersion. For all the simulations in the main text, we have used transfer matrix approach including dispersion. The advantage of the transfer matrix approach is for large systems (>10x10) where the loss inside the link resonators could become appreciable.

\section*{S2. Estimation of system parameters and numerical methods}

\begin{figure}
\includegraphics[width=\textwidth]{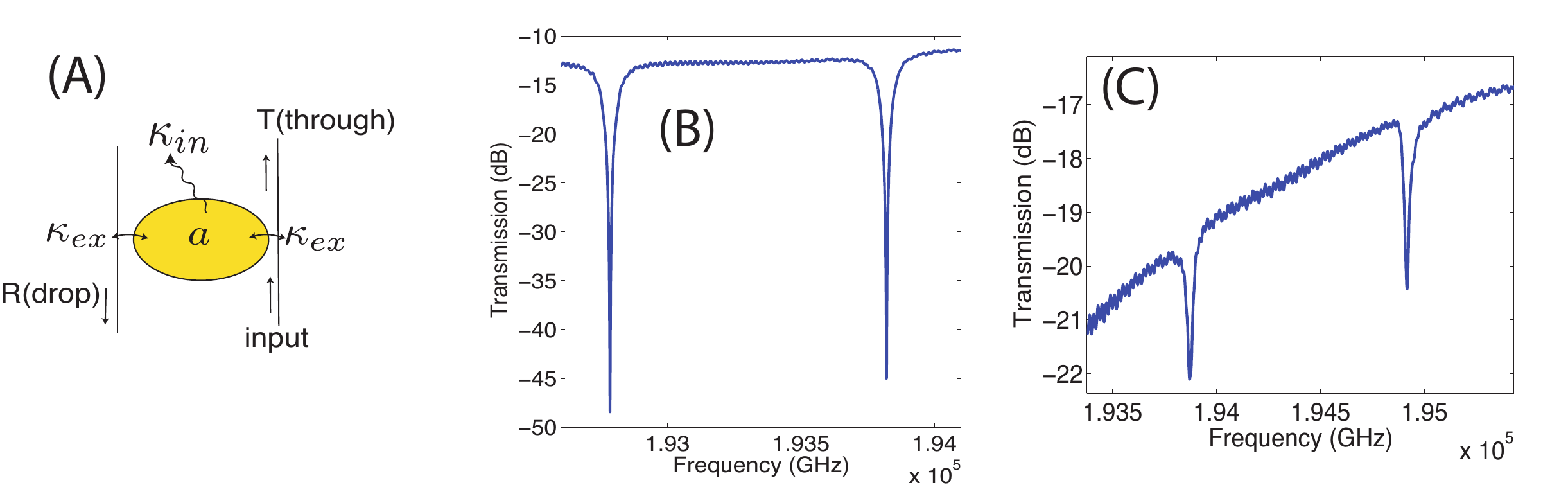}\caption{(a) An add/drop filter (b) Transmission spectrum in the through port for the add/drop filter (c) Transmission spectrum for a notch filter. \label{fig:Add/Drop-filter}}
\end{figure}

Here we show how different system parameters can be estimated using
simpler devices fabricated on the same chip. In particular, we consider an add/drop filter, as
shown in Fig.~\ref{fig:Add/Drop-filter}. Using coupled mode theory
(input-output formalism), we write the equation of motion of the resonator:

\begin{eqnarray*}
\dot{a} & = & (-\kappa_{\rm{in}}-2\kappa_{\rm{ex}})a+\sqrt{2\kappa_{\rm{ex}}}a_{\rm{in}}(t),
\end{eqnarray*}
where $a_{\rm{in}}$ is the input field and $\kappa_{\rm{ex}}(\kappa_{\rm{in}})$
are the extrinsic (intrinsic) decay rate of the field.
Assuming that the resonator is driven by a monochromatic field with
a detuning $\omega$ with respect to the resonator, we solve the system
in the steady-state. Consequently, the output transmission in the
drop and through channel are respectively given by:

\begin{eqnarray*}
R & =\left|\frac{\sqrt{2\kappa_{\rm{ex}}}a}{a_{\rm{in}}}\right|^{2} & =\frac{4\kappa_{\rm{ex}}^{2}}{\omega^{2}+(2\kappa_{\rm{ex}}+\kappa_{\rm{in}})^{2}}\\
T & =\left|\frac{a_{\rm{in}}+\sqrt{2\kappa_{\rm{ex}}}a}{a_{\rm{in}}}\right|^{2} & =\frac{\omega^{2}+\kappa_{\rm{in}}^{2}}{\omega^{2}+(2\kappa_{\rm{ex}}+\kappa_{\rm{in}})^{2}}.
\end{eqnarray*}
Therefore, in the through port, the full width half maximum (FWHM)=$2(\kappa_{\rm{ex}}+\kappa_{\rm{in}})$
and the contrast is $T_{\rm{max}}/T_{\rm{min}}=(2\kappa_{\rm{ex}}+\kappa_{\rm{in}})^{2}/\kappa_{\rm{in}}^{2}$.
Consequently, the measured quantities for the contrast and FWHM give
us both decay rates. Specifically, from the measurement we get $\kappa_{\rm{ex}}=$15 GHz, $\kappa_{\rm{in}}=$1 GHz.  We note that the parameters change from device to device and we have considered this disorder in our analysis, as we discuss below.

We have also obtained these numbers for the notch filter, where the
transmission in the single-mode approximation is given by:

\begin{equation}
T'=\frac{\omega^{2}+(\kappa_{\rm{ex}}-\kappa_{\rm{in}})^{2}}{\omega^{2}+(\kappa_{\rm{ex}}+\kappa_{\rm{in}})^{2}}.
\end{equation}
Again the contrast and FHWM yields both the intrinsic and extrinsic
decay rates $\kappa_{\rm{ex}}\approx$ 13 GHz, $\kappa_{\rm{in}}\approx$ 2 GHz. Note since we are
operating at the over-coupled regime $\kappa_{\rm{ex}}\gg\kappa_{\rm{in}}$,
the accuracy of the intrinsic loss is worse than that of the extrinsic
loss.

For the simulation of Fig.~2 of the main text, we estimate the tunneling rate from the
transmission of the 2D device, in particular, $6J\approx95$ GHz. We used
these parameters for the numerical simulation, $\kappa_{\rm{ex}}/J\approx0.8,\kappa_{\rm{in}}/J\approx0.06$. The tunneling rate can be also measured using a  three-ring device (Fig.\ref{fig:three-rings}). The uniform magnetic flux per plaquette is estimated to be $\alpha=2xn/\lambda\approx 0.15$, in the stripes, according to our design. No direct measurement of $\alpha$ could be made. More precisely, we used the following gauge: for the tunneling in the y-direction we chose the phase to be zero, and for the tunneling in the x-direction, we increased the phase by $\alpha$ as a function of row number mod 4: zero phase for the first row,  $\alpha$ for the second row, $2 \alpha$ for the third row and $3 \alpha$ for the forth row, zero for the five row etc. This leads to a uniform flux $\alpha$ per plaquette in the stripes of 10x4 and a $-3\alpha$ phase slip in the region connecting the stripes. We used this parameters and the  transfer matrix formalism to obtain $T_{12}$ and $T_{34}$ for a system without disorder.

We have different source of disorders which we characterize and estimate separately as different parameters of the generalized coupled mode theory (similar to Ref.~\cite{Hafezi:2011delay}).   We have assumed that the frequency mismatch has a gaussian distribution with a standard deviation $U=0.8 J$. The estimate is based on measuring six notch filters with identical design that were fabricated on a single chip. This result was consistent across several chips. Note that the chip-to-chip variation is usually higher by a couple of orders of magnitude, however, the variation within a chip, and more importantly, the variation within in a device on a  chip is lower.  We used measurements of single resonators to get an estimate of the variation in the tunneling rate ($J$) and the intrinsic loss rate ($\kappa_{\rm{in}}$). The parameters are: $\Delta \kappa_{\rm{in}}/\kappa_{\rm{in}}=0.45, \Delta J/J=0.04$. The magnetic phase variation is estimated to be  $\Delta \alpha=2 \pi$  0.08.

We have  used these parameters to obtain the variation between different disorder realizations and presented it as the gray band in Fig.~2B of the main text.  We have plotted $\pm$ 2 standard deviations as the confidence band. To obtain the dispersion relation (Fig.~2A in the main text), we follow Ref.~\cite{Hafezi:2011delay}.

Due to the roughness of the waveguide walls, the light can backscatter inside the resonator. The backscattering can be manifested as the mode coupling between the clockwise and the counter-clockwise modes of the resonator (i.e., spin flip), which in turn can be seen as a splitting in the  transmission signals. Such mode splitting  is not visible in our case, as shown in Fig.~\ref{fig:Add/Drop-filter}. We note that for an ultra high quality factor the aforementioned ratio increases and the mode coupling can be seen. To quantify the backscattering rate, we directly measured the backscattered signal ( $T_{14}$) for a single and three resonators and the rate is estimated to be $\beta/J \approx 0.04$.  We show the backscattering signal for a typical disordered system in Fig.~2B.

We find the eigenstates of the system and calculate the dispersion relation by assigning a momentum to each eigenstate which is chosen to be the relative phase between two resonators on the edge. Since edge states are semi-one dimensional, they have a well-defined momentum while such momentum assignment to bulk states leads to random phases. The edge and bulk bands are identified in Fig.~2A of the main text. The edge states route around the stripes as shown Fig.~2D of the main text. Similarly, for Figures 3,4, the simulation parameters are measured for single and three-ring devices and the transfer matrix approach was used to simulate a system without disorder.  The parameters are reported in the figure captions.

\section*{ S3. Experimental methods and Processing}

\begin{figure}
\includegraphics[width=0.4\textwidth]{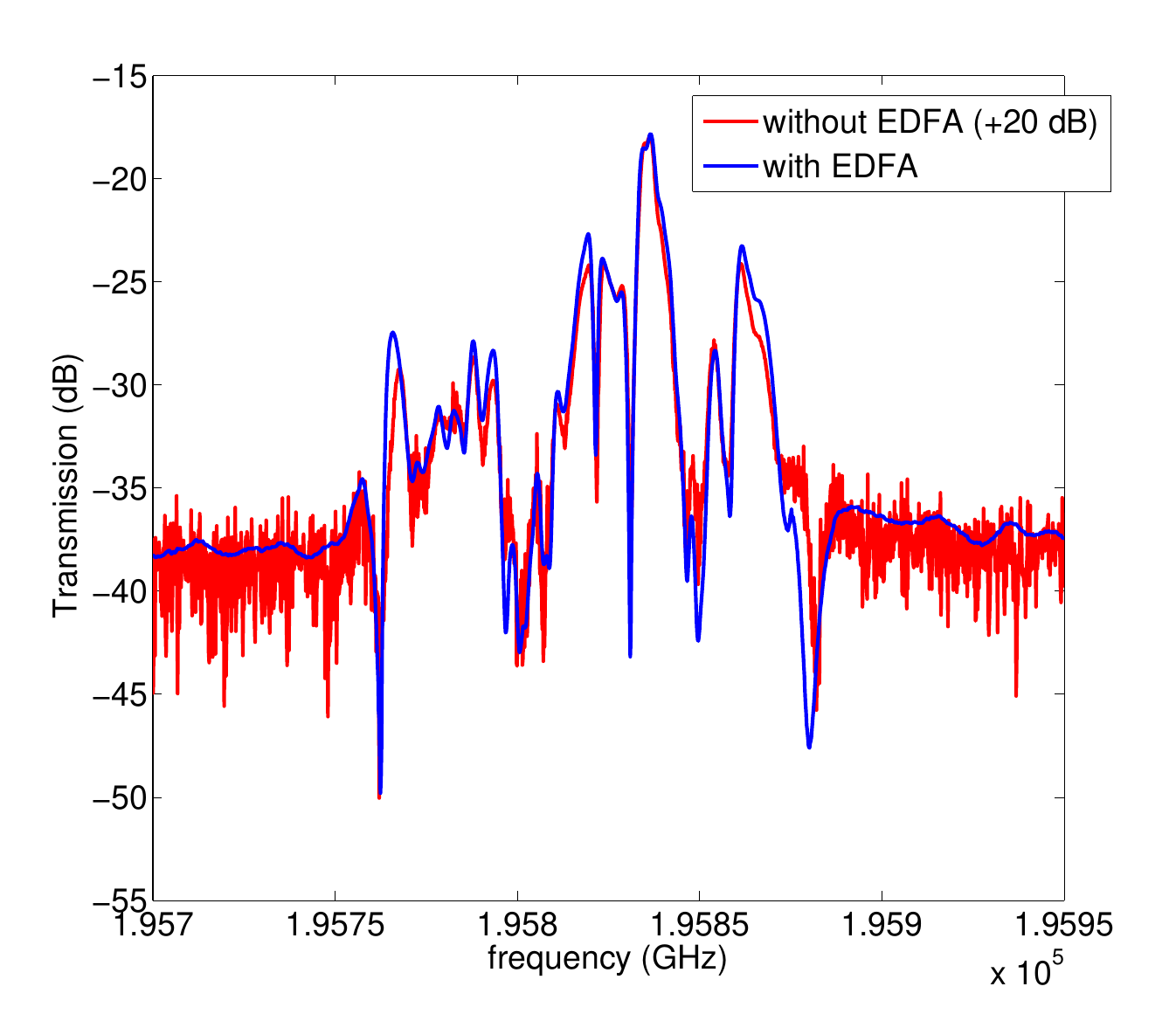}\caption{Transmission spectrum with and without EDFA for a 8x8 lattice\label{fig:EDFA}}
\end{figure}

We used straight and curved gratings to couple light into and out of the photonic chip. The efficiency of
such gratings was limited and estimated to be roughly 8 dB (3dB) for the straight (curved) gratings. The fiber connections efficiencies were estimated to be better than 1.5
dB. The remaining loss was due to the scattering of resonator's light
out of the guided mode, which varies depending on the number and the size of the resonators.
To have a significant signal to noise ratio, we amplified
the LUNA OVA source light using an  Erbium Doped Fiber Amplifier (EDFA) in the saturated regime with
constant output power + 17dBm, as shown in Fig.~1C in
the main text \cite{Cooper:2010ua}. Before sending the light to the
chip,  a tunable attenuator was used to adjust
the intensity of the light to avoid  nonlinear response
of the system. As shown in Fig.~\ref{fig:EDFA}, the amplification
does not change the spectrum. Polarizer paddles were used to maximize
the coupling to the TE mode of the waveguides and to increase the
signal to noise ratio for the IR imaging. The contrast seen by the camera
for the incoming photon polarization was better than 23 dB.

\begin{figure}
\includegraphics[width=0.7\textwidth]{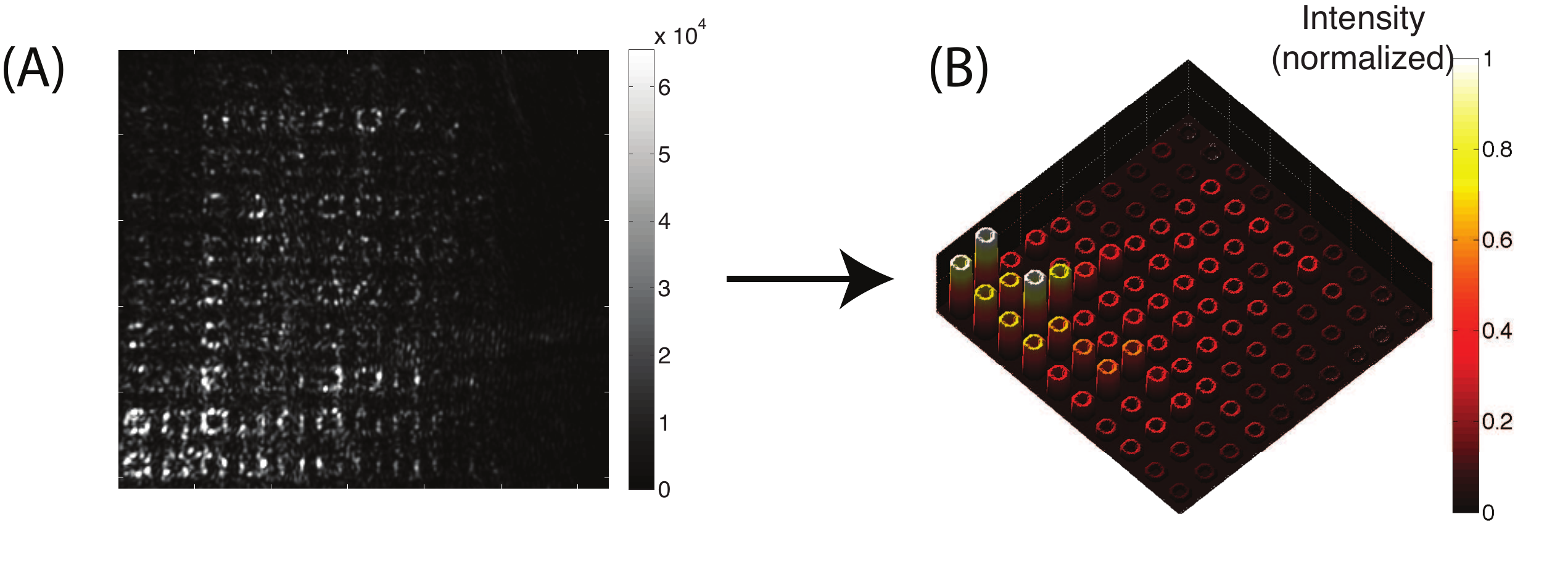}\caption{(a) Raw image from the camera.(b) Processed image after averaging
over 10x10 cells. \label{fig:image_processing}}
\end{figure}

Since the grating couplers do not have the same efficiency, we rescaled $T_{12}$ and $T_{34}$ for comparison by normalizing the averaged intensity over the entire band,  to generate Fig.~2C of the main text. To obtain the IR figures, such as Fig.~2E-G of the main text, we decreased the intensity of the input light
to avoid saturating the camera. To reduce the effect of background
light, e.g. due to scattering from the fiber holder, lens etc., and
to remove speckles, we used an averaging technique. More specifically,
the raw image was diced into 10x10 cells so that each cell contains
only one resonator and the signal was averaged over that cell (see
Fig.~\ref{fig:image_processing}). The averaged intensity was then
used to generate the 3D plot, where the height of each column represents
the intensity of the light circulating in each resonator. We used
a power-law mapping of the measured intensity for plotting.

Moreover, for Fig.~2 of the main text, an image was taken when the system was excited at a frequency out
of the transmission band. This image was used as the background noise
to be subtracted in our imaging process. For Figs. 3,4 of the main text, such subtraction was not required since the grating couplers were more efficient and the background light was not significant.

We note the imaged intensity of the bulk states could be sometimes higher or lower than the edge states. In fact, depending on the impurity configuration, some bulk state could exhibit higher transmission than the edge state. For the images in Figs. 2,3, we normalized the imaged to unity to make the features of the different states more apparent.

\newpage
\section*{ S4.  Resonator intensity measurement }

\begin{figure}
\includegraphics[width=0.5\textwidth]{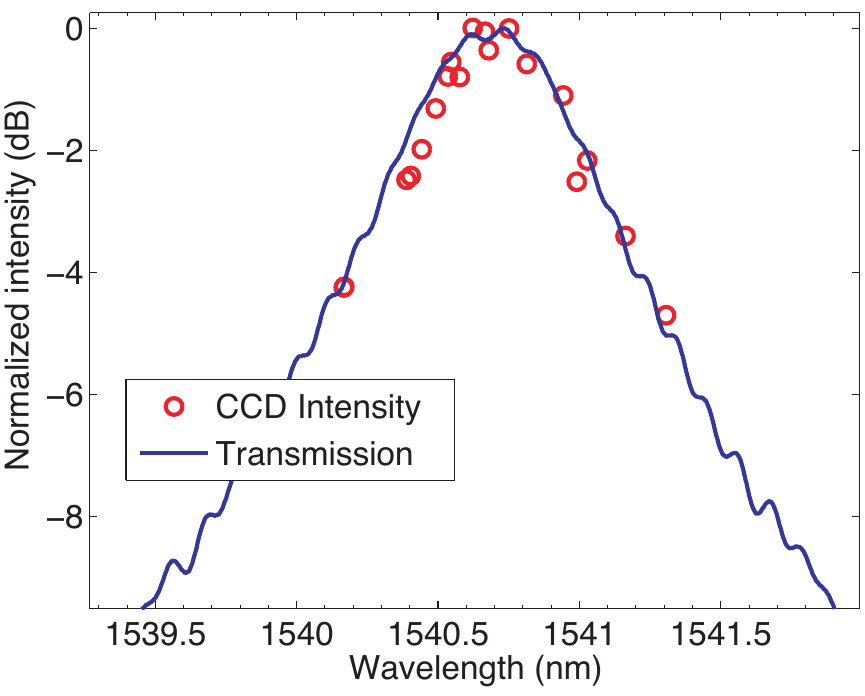}\caption{The measured  intensity of the rings using CCD camera is proportional to the output intensity in the drop port.\label{fig:CCD_camera_intensity}}
\end{figure}

To correlate the observed CCD Camera intensity to the ring resonator intensity, we used an add-drop filter (ADF). The ADF was fabricated on the same chip and with the same dimensions as the ring resonators in the 2D array devices. The transmission spectrum of the ADF was first measured at the drop port using the LUNA OVA 5000. The wavelength coupled to the ADF was then slowly scanned to get IR images at different wavelengths. The normalized transmission spectrum and the observed CCD camera intensity are shown in Fig.~\ref{fig:CCD_camera_intensity}. The CCD camera intensity shown here is the  intensity after integrating over the ring area.
Since the transmitted power at the drop port is proportional to the ring resonator intensity, this plot shows a very good correlation between the observed CCD camera intensity and the ring resonator intensity. Therefore, even if a direct measurement of the light intensity traveling in a ring resonator sitting in a 2D array is not possible, the CCD camera imaging allows an indirect relative measurement of the power distribution in ring resonators of the 2D array and hence facilitates the direct observation of edge states.
Note that the observed intensity in the CCD camera is a function of the light intensity traveling in the ring and also the loss factor $\kappa_{\rm{in}}$ of the ring. So if there are multiple resonators with equal light intensity traveling through them but with different $\kappa_{\rm{in}}$, the corresponding intensities read by the CCD camera imaging will be different. In our structures, since the variations in $\kappa_{\rm{in}}$ are of the order of 0.5 $\kappa_{\rm{in}}$, we expect some discrepancy between the simulated and the experimentally observed ring resonator intensities on the edge states. Such variation could be responsible for the discrepancy between the smooth edge state profile in the simulation and inhomogenous  profile of the edge states in the experimental images.

\newpage
\section*{ S5.  Transmission spectrum for the  device in Fig.~2 of the main text   }

\begin{figure}[h]
\includegraphics[width=0.8\textwidth]{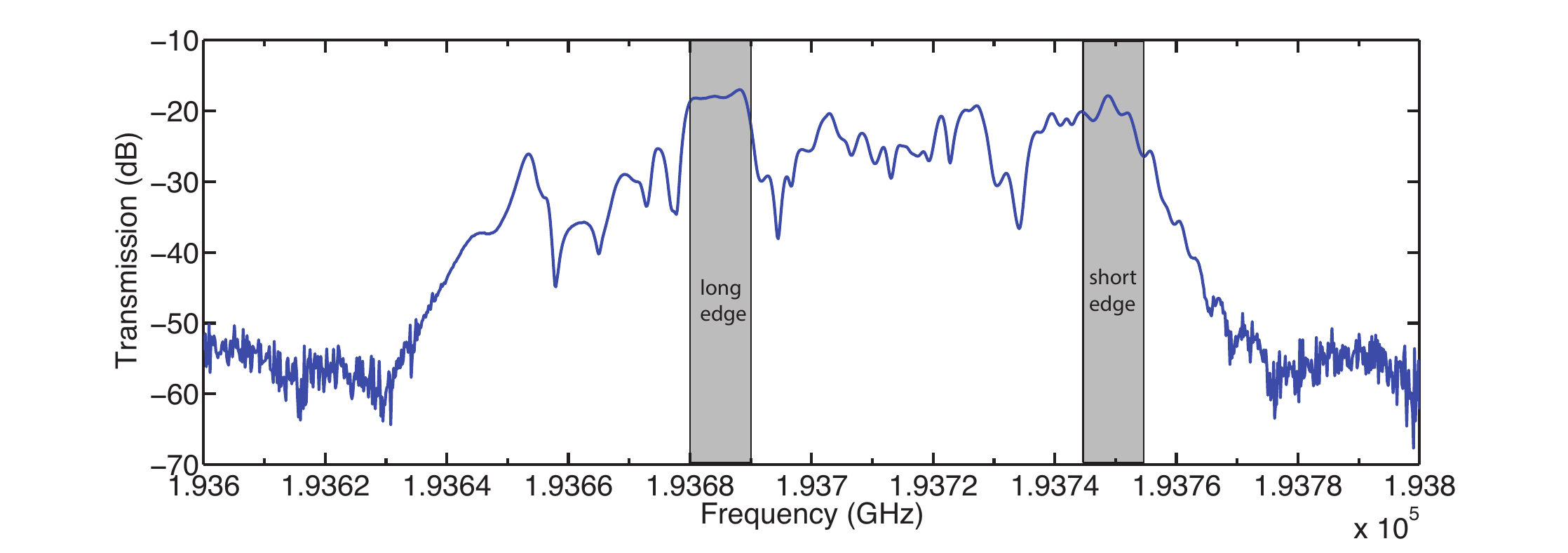}\caption{Transmission spectrum for the device used in Fig.~2 of the main text. The long and the short edge bands are highlighted in gray.\label{fig:8x8_spectrum}}
\end{figure}

\bibliographystyle{apsrev}